\documentclass[dvips,aps]{revtex4}
\usepackage{amsmath} 
\usepackage{amsfonts}
\usepackage{amstext}
\usepackage{latexsym, amssymb} 
\usepackage{bm} 
\usepackage{hyperref}
\begin{document}
\title{Quantum and classical features in the explanation of
  collisional decoherence}
\author{Bassano \surname{Vacchini}}
\email{bassano.vacchini@mi.infn.it}
\affiliation{Dipartimento di Fisica
dell'Universit\`a di Milano and INFN,
Sezione di Milano
\\
Via Celoria 16, I--20133, Milan, Italy}
\begin{abstract}
A most simple theoretical argument is given in order to explain the
quantitative estimate of the effect of collisional decoherence in
matter-wave interferometry. The argument highlights the relevance of
quantum and classical features in the description of the phenomenon,
showing in particular the connection between the formula used for the
experimental fit and the loss term in the classical linear Boltzmann equation.
\end{abstract}
\maketitle

\section{Introduction}
\label{sec:introduction}

Paradoxical as it may seem, one of the most intriguing features of the
quantum world is certainly its relationship to the classical
one~\cite{KieferNEW}. Apart from the conceptual charm this question is
of practical relevance in many respects. Full understanding of this
issue would give an answer to which quantum phenomena might appear
or persist on a meso- or macroscopic scale, to the degree of
feasibility of quantum computation, to the reliance of classical
pictures and classical computations, often simpler and more intuitive.
First significant steps in this direction are certainly given by
recent experiments in which loss of coherence due to the interaction
with the environment can be quantitatively studied. This phenomenon
goes under the name of decoherence, and its theoretical description is
based on an understanding of open system dynamics~\cite{Petruccione}.
\par
In this brief paper we will focus on recent beautiful experiments
concerned with the observation of collisional decoherence in
matter-wave interferometry~\cite{ZeilingerQBM-exp}, sketchily giving a
most simple theoretical argument for the loss of visibility, thus
showing the connection with the classical linear Boltzmann equation
and further giving an exact expression for the macroscopic scattering
cross-section.

\section{Analysis of collisional decoherence}
\label{sec:analys-coll-decoh}

In the aforementioned experiments the visibility of the interference
fringes, obtained letting a beam of fullerenes go
through a Talbot-Laue interferometer, is progressively reduced by
raising the pressure of the background gas present in the experimental
apparatus. The visibility decreases according to the formula
\begin{equation}
   \label{eq:1}
   V=V_0 \exp (-\Gamma t)
\end{equation}
where $V_0$ is the reference value, $t$ the time of flight through the
apparatus, $\Gamma$ a decoherence rate given by
\begin{equation}
   \label{eq:2}
   \Gamma=n v_0 \sigma_{\scriptscriptstyle \mathrm{macro}} (v_0)
\end{equation}
with $n$ the density of the gas, $M$ the mass of the fullerenes,
$v_0=p_0/M$ the modulus of their incoming velocity, supposed to be
essentially unchanged despite the collisions with the background
particles, $\sigma_{\scriptscriptstyle \mathrm{macro}} (v_0)$ the
total macroscopic scattering cross-section off the background gas,
parametrically depending on the modulus of the incoming velocity or
equivalently on momentum. To give a theoretical justification of the
formula~(\ref{eq:1}) besides a detailed characterization of the
experimental apparatus (in this connection see
also~\cite{garda03post}) one essentially needs a quantum mechanical
description of the fullerene dynamics due to collisions with a
surrounding gas. This is achieved by means of a quantum
master-equation which, as the underlying physics in terms of
collisions obviously tells, is actually a quantum counterpart of the
classical linear Boltzmann equation. The history of this
master-equation is relatively long, and we refer the reader
to~\cite{vienna} for a discussion of this point. A general result in
this direction, obviously still open to improvements is given
in~\cite{art7}, to which we refer for further detail and
references. The master-equation takes the form
\begin{eqnarray}
  \label{nalbe}
  &&   
  {  
        d {\hat \rho}  
        \over  
                      dt
        }  
        =
        -
        {i \over \hbar}
        [
        {\hat {{\sf H}}}_0
        ,
        {\hat \rho}
        ]
                +
        {n \over M^2}
        \int d^3\!
        {{\bm{q}}}
        \,  
        \Sigma (q)
  \\
 &&
 \hphantom{ju}
 \times
      \Biggl[
        e^{{i\over\hbar}{{\bm{q}}}\cdot{\hat {{\sf x}}}}
        \sqrt{
        S({{\bm{q}}},{\hat {{\sf p}}})
        }
        {\hat \rho}
        \sqrt{
        S({{\bm{q}}},{\hat {{\sf p}}})
        }
        e^{-{i\over\hbar}{{\bm{q}}}\cdot{\hat {{\sf x}}}}
        -
        \frac 12
        \left \{
        S({{\bm{q}}},{\hat {{\sf p}}}),
        {\hat \rho}
        \right \}
        \Biggr],
        \nonumber
\end{eqnarray}
with ${\hat {{\sf x}}}$ and ${\hat {{\sf p}}}$ position and momentum
operator for the test particle (presently the fullerene), ${\hat {{\sf
      H}}}_0={\hat {{\sf p}}}^2 / 2M$, $\Sigma (q)=M^2
(2\pi\hbar)^3\frac{2\pi}{\hbar}| \tilde{t} (q) |^2$ ($\tilde{t} (q)$
being the Fourier transform of the interaction potential),
$S({{\bm{q}}},{\hat {{\sf p}}})$ the operator-valued dynamic structure
factor, given by the Fourier transform of the time dependent spatial
autocorrelation function of the gas
\begin{eqnarray}
   \label{eq:3}
  {S} ({\bm{q}},E ({\bm{q}},{\bm{p}}))&=&
        {  
        1  
        \over  
         2\pi\hbar
        }  
        \int dt 
        {\int d^3 \! {\bm{x}} \,}        
        e^{
        {
        i
        \over
         \hbar
        }
        (E({\bm{q}},{\bm{p}}) t -
        {\bm{q}}\cdot{\bm{x}})
        }  
      \\
\nonumber
&&
\hphantom{        {  
        1  
        \over  
         2\pi\hbar
        }  
        \int dt 
}
\times
        {
        1
        \over
         N
        }
        {\int d^3 \! {\bm{y}} \,}
        \left \langle  
         N({\bm{y}})  
         N({\bm{x}}+{\bm{y}},t)
         \right \rangle.
\end{eqnarray}
with ${\bm{q}}$ and $E\equiv E({\bm{q}},{\bm{p}})={ q^2 \over 2M } + {
  {\bm{p}} \cdot {\bm{q}} \over M }$ momentum and energy transferred
to the fullerene molecule in each single collision. The two-point
correlation function ${S} ({\bm{q}},E)$ is related as shown by van
Hove~\cite{vanHove} to the macroscopic differential scattering
cross-section of a test particle off a medium through the formula
\begin{equation}
  \label{diff}
  \frac{d^2 \sigma_{\scriptscriptstyle \mathrm{macro}}}{d\Omega dE}  =
\frac{p'}{p}\Sigma (q)
  S ({\bm{q}},E).
\end{equation}
Eq.~(\ref{nalbe}) correctly avoids the extra $2\pi$
factor appearing for example in the well-known result of Gallis and
Fleming~\cite{Gallis90}, as stressed in~\cite{ZeilingerQBM-th} and
also found in~\cite{HalliwellQBM}, in both of which a particular example
of dynamic structure factor also implicitly appears. The
redundancy of this $2\pi$ factor can be best understood in physical
terms as in~\cite{Petruccione}, where the authors already stressed
that this factor would lead to a decoherence rate given by $2\pi$ times the
total scattering rate. Experiments do in fact rule out this factor.
With respect to other results the quantum linear Boltzmann equation
(\ref{nalbe}) encompasses the momentum of the test particle as a
dynamical variable, thus allowing for a correct description of energy
transfer and for the expected thermal stationary solution~\cite{art5}.
Since on the time scale of the experiment thermalization does not play
a role, as a first simplification the operator ${\hat {{\sf p}}}$
in $S({{\bm{q}}},{\hat {{\sf p}}})$  can be replaced by ${\bm{p}}_0$, the
incoming momentum of the fullerene. Eq.~(\ref{nalbe}) thus becomes,
also exploiting~(\ref{diff})
\begin{equation}
   \label{eq:4}
  {  
        d {\hat \rho}  
        \over  
                      dt
        }  
        =
        -
        {i \over \hbar}
        [
        {\hat {{\sf H}}}_0
        ,
        {\hat \rho}
        ]
        +
        {n \over M^2}
        \int d^3\!
        {{\bm{q}}}
        \,  
        \Sigma (q)S({{\bm{q}}},{\bm{p}}_0)
        e^{{i\over\hbar}{{\bm{q}}}\cdot{\hat {{\sf x}}}}
        {\hat \rho}
        e^{-{i\over\hbar}{{\bm{q}}}\cdot{\hat {{\sf x}}}}
        -
        n\frac{p_0}{M}\sigma_{\scriptscriptstyle \mathrm{macro}} (p_0)
        {\hat \rho}.
\end{equation}
Note that the dependence on the incoming momentum of the test particle is still
crucial in order to explain the velocity dependence in the macroscopic
scattering cross-section appearing in~(\ref{eq:2}). The quantum nature
of~(\ref{eq:4}), apart from the free evolution responsible for the
interference phenomena, only appears in the second term at the r.h.s.,
the loss term simply being a multiplicative factor as in the classical
linear Boltzmann equation, apart from the obvious fact that the
microscopic scattering cross-section describing the single collision
is based on a quantum or at least a semi-classical
calculation~\cite{Williams}. The second term at the r.h.s.
of~(\ref{eq:4}) however does not contribute to the interference
fringes, because of its rapid oscillations leading to a destruction of
the off-diagonal matrix elements as described e.g.
in~\cite{AlickiPRA,Petruccione}.  Therefore eq.~(\ref{eq:4}) predicts
an exponential decay of the intensity relevant for the interference
fringes according to the rate
$n\frac{p_0}{M}\sigma_{\scriptscriptstyle \mathrm{macro}} (p_0)$,
which can be simply calculated from the loss term in the classical
linear Boltzmann equation. Following~\cite{Williams}, and expressing
the scattering cross-section as a function of the incoming velocity, we
can write
\begin{equation}
   \label{eq:5}
   n v_0\sigma_{\scriptscriptstyle \mathrm{macro}} ({\bm{v}}_0)=nv_0\int
   d^3\!u\, f_{\scriptscriptstyle \mathrm{MB}}
   ({\bm{u}})\frac{|{\bm{v}}_0-{\bm{u}}|}{v_0}
   \sigma_{\scriptscriptstyle \mathrm{micro}} ({\bm{v}}_0-{\bm{u}}) ,
\end{equation}
where $f_{\scriptscriptstyle \mathrm{MB}} ({\bm{u}})$ is the thermal
Maxwell-Boltzmann distribution of velocities in the background gas.
For a microscopic scattering cross-section of the form
\begin{equation}
   \label{eq:6}
   \sigma_{\scriptscriptstyle \mathrm{micro}}
   ({\bm{v}}_0-{\bm{u}}) =K |{\bm{v}}_0-{\bm{u}}|^{\alpha}
\end{equation}
the integral in~(\ref{eq:5}) can be exactly calculated.  Introducing
the most probable velocity ${v_{\scriptscriptstyle
    \mathrm{mp}}}=\sqrt{2/\beta m}$ of gas particles with temperature
$1/\beta$ and mass $m$ one has
\begin{align}
   \label{eq:7}
   \sigma_{\scriptscriptstyle \mathrm{macro}} (v_0)&=
   \frac{K}{v_0}\int
   d^3\!u\, u^{\alpha+1}f_{\scriptscriptstyle
     \mathrm{MB}}({\bm{u}}+{\bm{v}}_0)
\\ 
\nonumber
&=
K
\frac{2}{\sqrt{\pi}} 
\frac{v^{\alpha+2}_{\scriptscriptstyle \mathrm{mp}}}{v_0}
e^{-\left(
    \frac{v_0}{v_{\scriptscriptstyle \mathrm{mp}}}\right)^2}
\int_0^{+\infty}dt\,
t^{\alpha+2}e^{-t^2}\sinh\left(2t\frac{v_0}{v_{\scriptscriptstyle
      \mathrm{mp}}} \right) 
\end{align}
and exploiting the result~\cite{Ryzhik}
\begin{displaymath}
   \int_0^{+\infty} dx\, x^{2\mu-1}e^{-x^2}\sinh (\gamma
   x)=\frac{\gamma}{2}\Gamma \left(\mu+\frac{1}{2}\right)
e^{\frac{1}{4}\gamma^2}
   \Phi\left(1-\mu,\frac{3}{2};-\frac{1}{4}\gamma^2\right) 
\end{displaymath}
valid for $\mu>-\frac{1}{2}$, with $\Phi (\alpha,\beta;\gamma)$ the
confluent hypergeometric function, one immediately obtains the exact
expression
\begin{equation}
   \label{eq:8}
   \sigma_{\scriptscriptstyle \mathrm{macro}} (v_0)=K
\frac{2}{\sqrt{\pi}}\Gamma
\left(\frac{\alpha}{2}+2\right)\frac{v^{\alpha+1}_{\scriptscriptstyle
    \mathrm{mp}}}{v_0} 
\Phi \left(-\left(\frac{\alpha}{2}+\frac{1}{2}\right),\frac{3}{2};-\left(
    \frac{v_0}{v_{\scriptscriptstyle \mathrm{mp}}}\right)^2\right).
\end{equation}
For the case of interest $\sigma_{\scriptscriptstyle \mathrm{micro}}$,
calculated on the basis of an isotropic potential $U (r)=-\frac{C_6}{r^6}$,
corresponds to the particular choice~\cite{ZeilingerQBM-exp,Maitland}
\begin{displaymath}
   K=\frac{\left(\pi^6\frac{3}{8}\right)^\frac{2}{5}}{\sin\left(\frac{\pi}{5}\right)\Gamma\left(\frac{2}{5}\right)}   \left(\frac{C_6}{\hbar}\right)^\frac{2}{5}
\qquad \qquad\alpha=-\frac{2}{5}
\end{displaymath}
thus recovering, when considering only the first contributions in the expansion
in terms of the small parameter $v_0/v_{\scriptscriptstyle
    \mathrm{mp}}$, the result given in~\cite{ZeilingerQBM-th1}
  \begin{equation}
     \label{eq:9}
     \sigma_{\scriptscriptstyle \mathrm{macro}}(p_0)=\frac{\left(\pi^6\frac{3}{8}\right)^\frac{2}{5}}{\sin\left(\frac{\pi}{5}\right)\Gamma\left(\frac{2}{5}\right)}   
\left(\frac{C_6}{\hbar}\right)^\frac{2}{5}
\frac{2}{\sqrt{\pi}}\Gamma
\left(\frac{9}{5}\right)\frac{v^{\frac{3}{5}}_{\scriptscriptstyle
    \mathrm{mp}}}{v_0} 
\left[
1+\frac{1}{5} \left(\frac{v_0}{v_{\scriptscriptstyle
      \mathrm{mp}}}\right)^2+
\mathcal{O}\left(\left(\frac{v_0}{v_{\scriptscriptstyle \mathrm{mp}}}\right)^4\right)\right].
\end{equation}
A full theoretical explanation of the observed data, also keeping the
detailed features of the interferometer into account, has been hinted
in~\cite{ZeilingerQBM-exp,ZeilingerQBM-th1} and will appear in a
further publication by the same group~\cite{ZeilingerQBM-th2}.  The
argument presented here, however, gives a remarkably simple
explanation of the experimental results by impinging on both quantum
and classical aspects.  On the one hand one needs a true quantum
linear Boltzmann equation in order to understand how the gain term in
the equation does not contribute to the interference pattern, due to a
suppression of off-diagonal terms, on the other hand the loss term
which quantitatively describes the reduction of visibility when
treated classically leads to the formula~(\ref{eq:5}) used to fit the
experimental data.
The main difference between the approach presented here and the one
outlined in~\cite{ZeilingerQBM-exp,ZeilingerQBM-th1} lies in the fact
that here we start with a quantum linear Boltzmann equation already
containing, through the dynamic structure factor and the dependence on
the momentum of the incoming particle, all the informations about the
macroscopic scattering cross-section relevant for the loss of
visibility. When considering a master-equation like the one by Gallis and
Fleming, instead, the macroscopic or effective scattering
cross-section depending on the velocity of the incoming fullerene has
to be introduced as an additional information.

\begin{acknowledgments}
   The author would like to thank Prof.~L.~Lanz and Dr.~K.~Hornberger
   for useful discussions. The work was supported by MIUR under
   Cofinanziamento and FIRB.
\end{acknowledgments}

\end{document}